\documentclass{icrc2009}

\usepackage{graphicx}   
\usepackage[caption=false]{caption}    
\usepackage[font=footnotesize]{subfig} 
\usepackage{fixltx2e}
\usepackage{url}

\newcommand{\shorttitle}[1]%
{\markboth{Proceedings of the 31\MakeLowercase{$^{st}$} ICRC, {\L}\'{o}d\'{z} 2009}{#1} }
\newcommand{\etal}{\MakeLowercase{\textit{et al. }}} 

\begin{document}
\title{Highlights of Recent Multiwavelength Observations \\
       of VHE Blazars with VERITAS}

\author{\IEEEauthorblockN{J. Grube\IEEEauthorrefmark{1} for the VERITAS collaboration\IEEEauthorrefmark{2}}
                            \\
\IEEEauthorblockA{\IEEEauthorrefmark{1}School of Physics, University College Dublin, Belfield, Dublin 4, Ireland}
\IEEEauthorblockA{\IEEEauthorrefmark{2}see R.A. Ong et al. (these proceedings) or http://veritas.sao.arizona.edu/conferences/authors?icrc2009}
}
\shorttitle{J. Grube \etal Multiwavelength Observations of VHE Blazars with VERITAS}
\maketitle

\begin{abstract}
We present long-term observations of several VHE (E $>$ 100 GeV) blazars 
with VERITAS, together with contemporaneous \emph{Swift} and \emph{RXTE} 
X-ray data. The observed targets include Mrk\,421, Mrk\,501, 
1ES\,2344+514. Strong flux and spectral variability is seen in Mrk\,421 
on nightly time-scales between January and June 2008, revealing a highly 
correlated X-ray to GeV/TeV connection. Modest X-ray variability is evident 
in Mrk\,501. Observations of 1ES\,2344+514 in December 2007 show 
VHE $\gamma$-ray and X-ray flux doubling on nightly time-scales.
\end{abstract}

\section{Introduction}
Blazars (BL Lac objects and Flat Spectrum Radio Quasars) are active galactic 
nuclei (AGN) with a relativistic plasma jet oriented close to the line of 
sight \cite{Blandford79}. These objects exhibit rapid variability and have 
broadband spectral energy distributions (SEDs) characterized, in 
a $\nu\rm{F}_{\nu}$ representation, by a synchrotron component extending from 
radio to X-ray frequencies, and a second component peaking at $\gamma$-ray 
frequencies due to either inverse-Compton radiation or from hadronic 
processes. BL Lac type blazars are further sub-divided based on the peak 
frequency of the synchrotron emission as low-, intermediate-, or 
high-frequency-peaked BL Lacs (LBLs, IBLs, and HBLs). Currently, 20 HBLs 
from a total of $\sim$25 VHE blazars are detected at very high energy 
(VHE, E $>$100 GeV) $\gamma$-rays.\footnote{TeVCat catalog of VHE 
$\gamma$-ray sources: http://tevcat.uchicago.edu} Elsewhere in these 
proceedings, the VERITAS blazar observing program is described 
\cite{Benbow09}, and the discovery of VHE emission with VERITAS from two 
HBLs (RGB\,J0710+591 and 1ES\,0806+524), and from two IBLs (3C\,66A and 
W\,Comae) is highlighted \cite{Perkins09}. 

Multiwavelength campaigns on BL Lacs are important to sample the broadband 
flux and spectral variability on time-scales ranging from minutes to 
months. Recent joint observing campaigns with VERITAS on two IBLs are 
presented elsewhere in these proceedings. Light curves and the broadband 
SED of 3C\,66A are presented from contemporaneous VERITAS, \emph{Fermi}, 
X-ray, and optical data between September and November 2008 \cite{Reyes09}. 
Observations of W Comae in June 2008 reveal X-ray and VHE $\gamma$-ray 
flaring, with a marginal detection above 100 MeV from AGILE observations 
\cite{Maier09}. In this paper detailed multiwavelength results on the 
bright and well-studied HBLs: Mrk\,421, Mrk\,501, and 1ES\,2344+514 are 
presented. Multiwavelength results on the recently observed HBL 
RGB\,J0710+591 will also be presented at the conference.
\section{VERITAS and X-ray Data Analysis}
\begin{figure}
\centering
\includegraphics[width=3.0in]{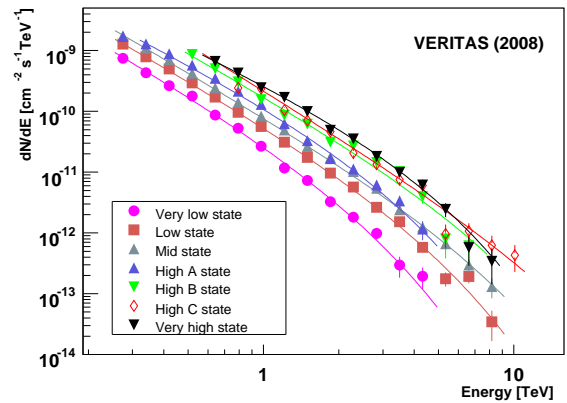}
\caption{Time-averaged VERITAS photon spectra of Mrk\,421 for discrete 
flux levels. A power law with exponential cutoff model $\rm{dN}/\rm{dE} = 
\rm{I}_{\rm{o}} \cdot (\rm{E}/1 \rm{TeV})^{-\Gamma} \cdot 
exp(\rm{-E}/E_{\rm{cut}})$ is fit to each spectrum.}
\label{FigMrk421spec}
\end{figure}
\begin{figure*}
\centering
\includegraphics[width=5.9in]{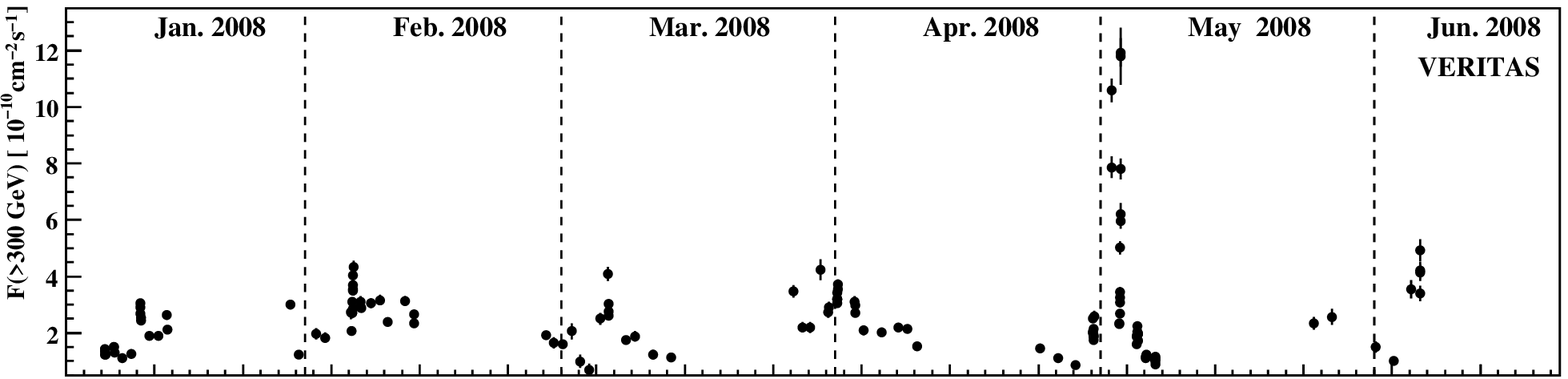} 
\includegraphics[width=5.9in]{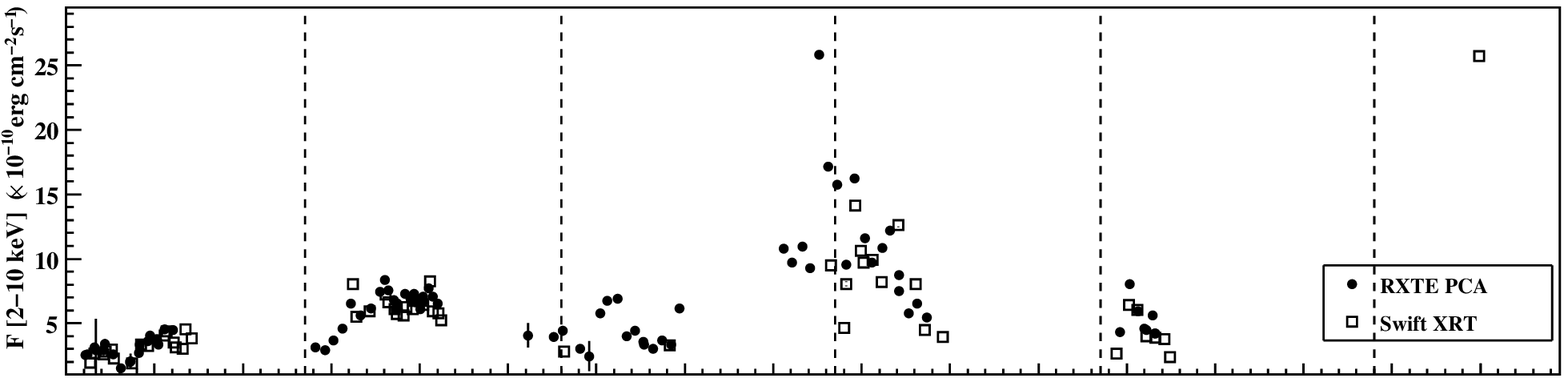}
\includegraphics[width=5.9in]{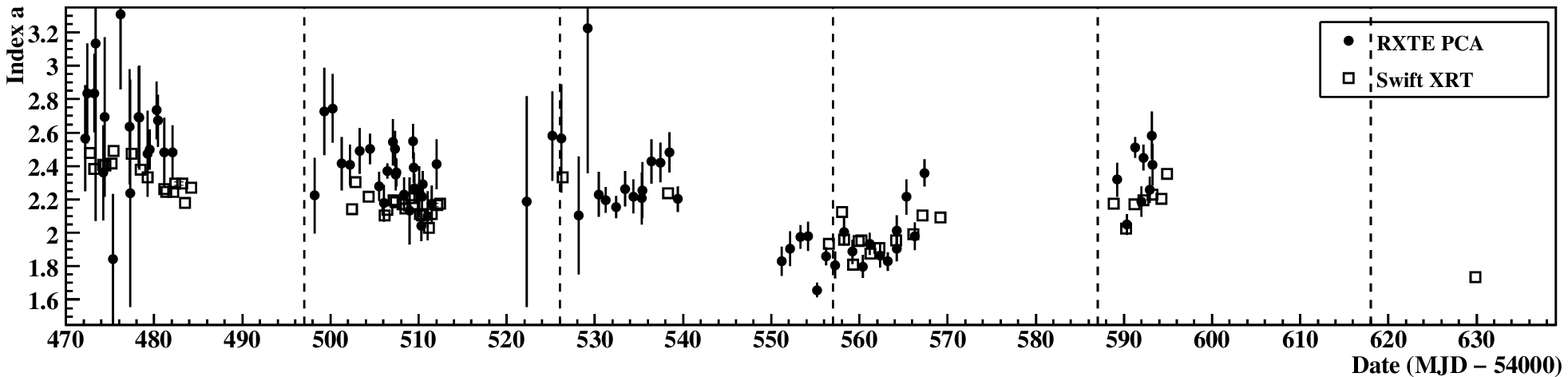}
\caption{VHE $\gamma$-ray and X-ray light curve of Mrk\,421. Shown 
in the top panel are VERITAS F($>$300 GeV) fluxes from observations of 
20 minutes each. The middle panel shows the 2--10\,keV fluxes from 
observations with \emph{RXTE} PCA (circles) and \emph{Swift} XRT 
(open squares). The bottom panel shows the X-ray spectral indices ($a$ 
parameter, see text) from log-parabolic fits to the 0.4--10\,keV \emph{Swift} 
XRT spectra and 3--20\,keV \emph{RXTE} PCA spectra.} 
\label{FigMrk421curve}
\end{figure*}
VERITAS is an array of four imaging atmospheric-Cherenkov telescopes located 
in southern Arizona. The array is sensitive over the energy range of 
100 GeV to $>$30 TeV, and can detect (5 $\sigma$ level) a source flux of 
5\% of the Crab Nebula flux in $\sim$2.5 hours. All observations 
presented here pass run quality selection criteria, which removes data 
taken during poor weather conditions or non-standard hardware configurations. 
The \emph{wobble} mode, where the source is positioned at a fixed offset 
of 0.5$^{\circ}$ from the camera center, was used for all observations 
presented here. Standard data reduction and $\gamma$-ray selection cuts 
are applied \cite{Acciari08}. The results 
presented here agree well with those performed using an independent 
VERITAS analysis package.

Long-term X-ray observations with \emph{RXTE} PCA \cite{Jahoda96} and 
\emph{Swift} XRT \cite{Burrows05} were taken contemporaneously with the 
VERITAS data. All \emph{RXTE} PCA data were taken with only PCU2 operational. 
For all \emph{Swift} XRT data on Mrk\,421 the observations were taken in 
window timing (WT) mode, while for 1ES\,2344+514 the XRT data were all 
taken in photon counting (PC) mode. Data reduction is performed with 
the \emph{HEAsoft} 6.5 package, following the standard methods 
\cite{Grube08,Horan09}. In particular, annular source regions are used 
to extract the \emph{Swift} XRT data in PC mode when photon pileup is 
evident \cite{Grube08}.
\section{Mrk\,421}
\begin{figure}
\centering
\includegraphics[width=3.0in]{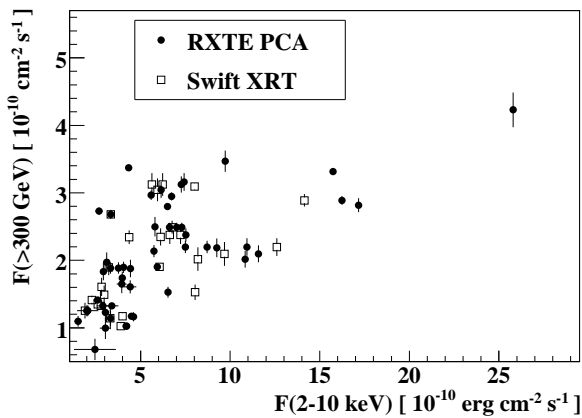}
\caption{VERITAS flux F($>$300 GeV) from 20 min exposures versus the 
X-ray flux for observations of Mrk\,421 taken within $\pm$5 hours.}
\label{FigMrk421Fcorr}
\end{figure}
\begin{figure*}
\centering
\includegraphics[width=5.5in]{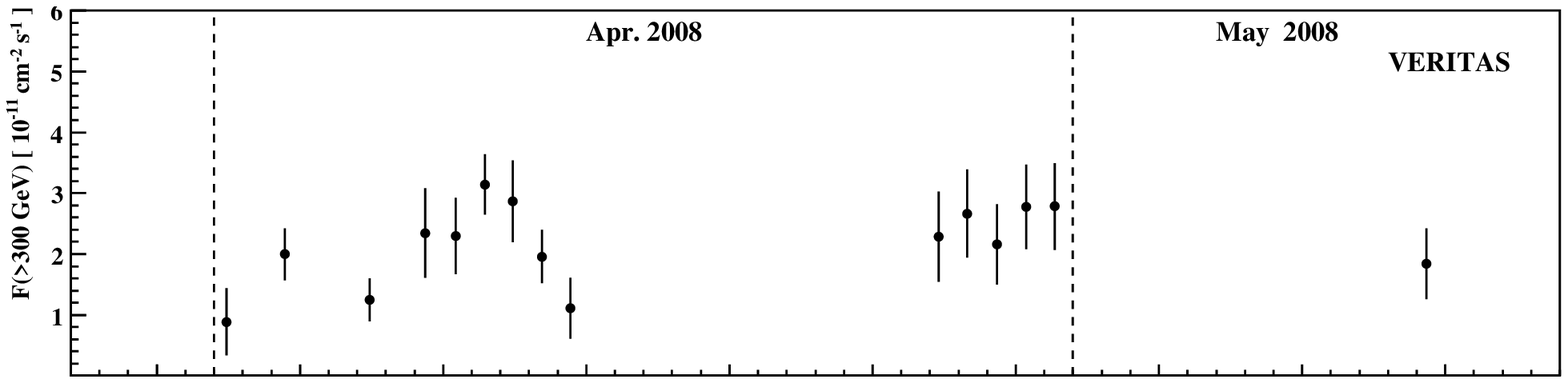}
\includegraphics[width=5.5in]{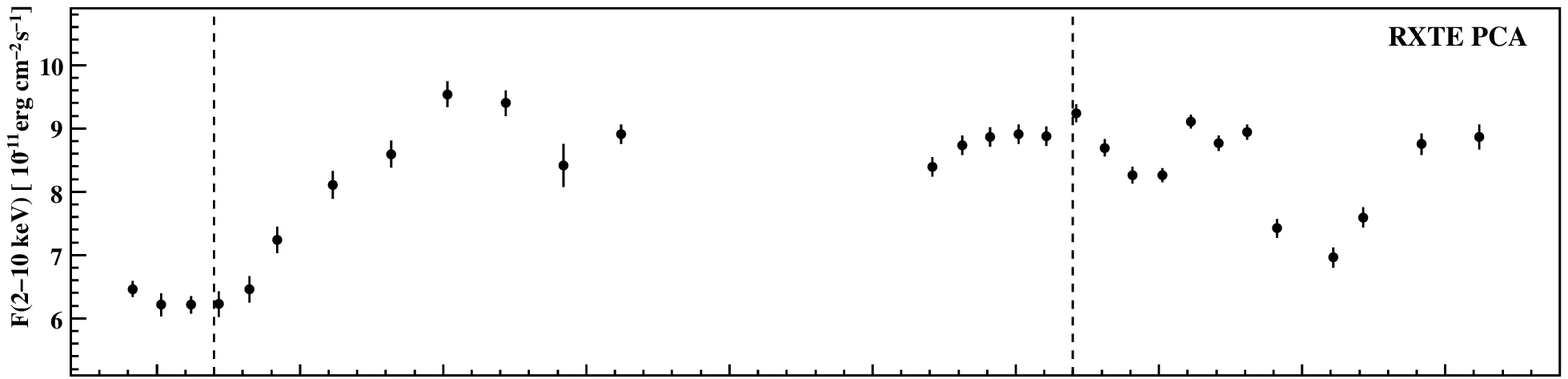}
\includegraphics[width=5.5in]{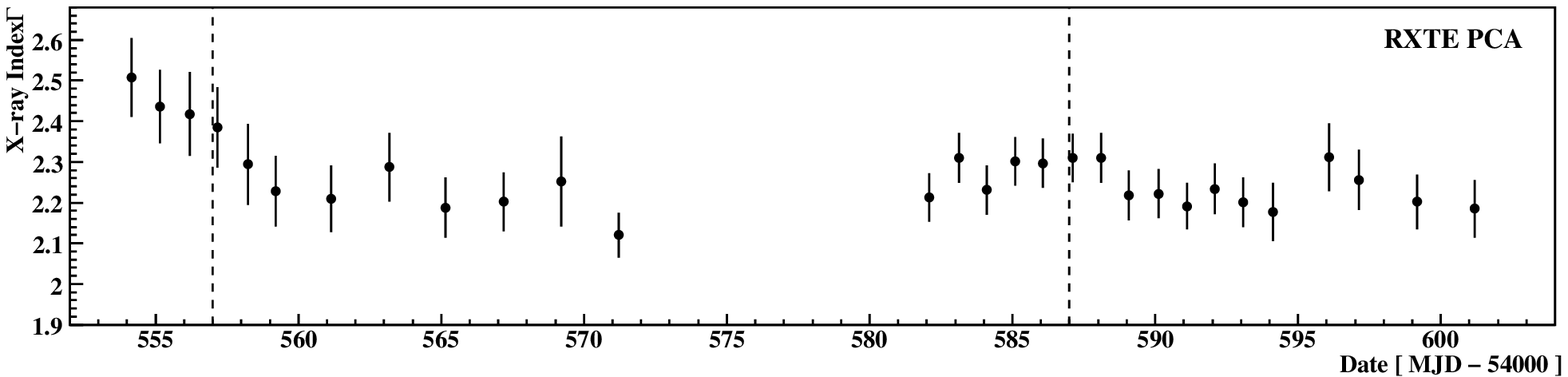}
\caption{VHE $\gamma$-ray and X-ray light curve of Mrk\,501. Shown 
in the top panel are the nightly VERITAS F($>$300 GeV) fluxes. The middle 
and bottom panels shows the \emph{RXTE} PCA 2--10\,keV fluxes and power law 
indices, respectively.
}
\label{FigMrk501curve}
\end{figure*}
Mrk\,421 is a nearby (z $=$ 0.031) HBL and is highly variable at UV to 
VHE $\gamma$-ray energies. VERITAS observed Mrk\,421 between January and June 
2008 for a total live-time of 43.6 hours. An average analysis energy threshold 
of 260 GeV is determined for the observations that span a range of zenith 
angles from 6$^{\circ}$--56$^{\circ}$. In the whole data set an excess of 
$\sim$30000 $\gamma$-ray events is detected, corresponding to a significance 
of $>$270 $\sigma$. Previously observed VHE $\gamma$-ray spectral variability 
correlated with flux level \cite{Krennrich02} is investigated by dividing 
the VERITAS data into subsets according to flux level. Figure 
\ref{FigMrk421spec} shows the VERITAS spectra in six subsets with best-fit 
curves for a power law with exponential cutoff model 
of the form dN/dE $=$ 
I$_{\rm{o}} \cdot (\rm{E}/1 \rm{TeV})^{-\Gamma} \cdot exp(\rm{-E}/E_{\rm{cut}})$. 
The nightly X-ray 
spectra from \emph{RXTE} PCA and \emph{Swift} XRT data are best fit with an 
absorbed log-parabolic model, which uses a fixed column density \cite{Horan09} 
and an energy dependent photon index 
$\Gamma = a + b \cdot \rm{Log}(\rm{E}/\rm{E}_{\rm{o}}$). The mean reduced 
$\chi^{2}$ values from log-parabolic fits to the \emph{RXTE} PCA and 
\emph{Swift} XRT spectra are 0.87 and 1.20 compared to reduced $\chi^{2}$ 
values of 1.35 and 1.80 for an absorbed power law model. 

Figure \ref{FigMrk421curve} shows on the top panel the nightly VHE 
$\gamma$-ray fluxes F($>$300 GeV) from nightly VERITAS observations. 
Shown on the middle and bottom panels are X-ray fluxes F(2--10 keV) and 
spectral indices from \emph{RXTE} PCA and \emph{Swift} XRT observations. 
Strong flux variability on nightly time-scales is seen over the entire 
six month period in 2008, with exceptionally bright X-ray flaring in March 
to April, and  VHE $\gamma$-ray flaring in early May. A measure of the 
integrated level of flux variability is the fractional root-mean-square (rms) 
variability amplitude F$_{\rm{var}}$ \cite{Vaughan03}. Significant 
VHE $\gamma$-ray flux variability of F$_{\rm{var}} = (66.5 \pm 0.6$)\% and 
X-ray flux variability F$_{\rm{var}} = (57.4 \pm 0.3$)\% was observed. 
Figure \ref{FigMrk421Fcorr} shows the VHE $\gamma$-ray 
flux versus X-ray flux for observations taken within $\pm$5 hours of each 
other. The highest near-simultaneous X-ray fluxes are from the March to 
April period. Unfortunately, when VERITAS measured the highest 
VHE $\gamma$-ray fluxes on May 2 and May 3 there were no simultaneous 
\emph{RXTE} PCA or \emph{Swift} data. Combining all VHE $\gamma$-ray flux 
versus X-ray flux points, the correlation coefficient is $r = 0.62 \pm 0.07$. 
Clear X-ray spectral hardening with increasing 2--10 keV flux is shown in 
figure \ref{FigMrk421curve}, with the log-parabolic index parameter $a$ 
ranging from $\sim$1.8--2.8.

\section{Mrk\,501}
The nearby HBL Mrk\,501 (z $=$ 0.034) was discovered at VHE $\gamma$-ray 
energies by the Whipple 10 m telescope in 1996 \cite{Quinn96}. VERITAS 
observed Mrk\,501 between April and June 2008 for a total live-time of 
6.2 hours. An excess with the statistical significance of 22 $\sigma$ 
is measured for the total data set.  Figure \ref{FigMrk501curve} shows the 
nightly VERITAS VHE fluxes F($>$300 GeV) and \emph{RXTE} PCA 2--10 keV fluxes 
and power law indices. Marginal variability is seen in the 
VERITAS F($>$300 GeV) fluxes with F$_{\rm{var}} = (14 \pm 12$)\%, while 
moderate 2--10 keV flux variability is evident with 
F$_{\rm{var}} = (12.7 \pm 0.4$)\%. More detailed results from this campaign 
are presented elsewhere in these proceedings \cite{Kranich09}.
\section{1ES\,2344+514}
\begin{figure*}
\centering
\includegraphics[width=5.9in]{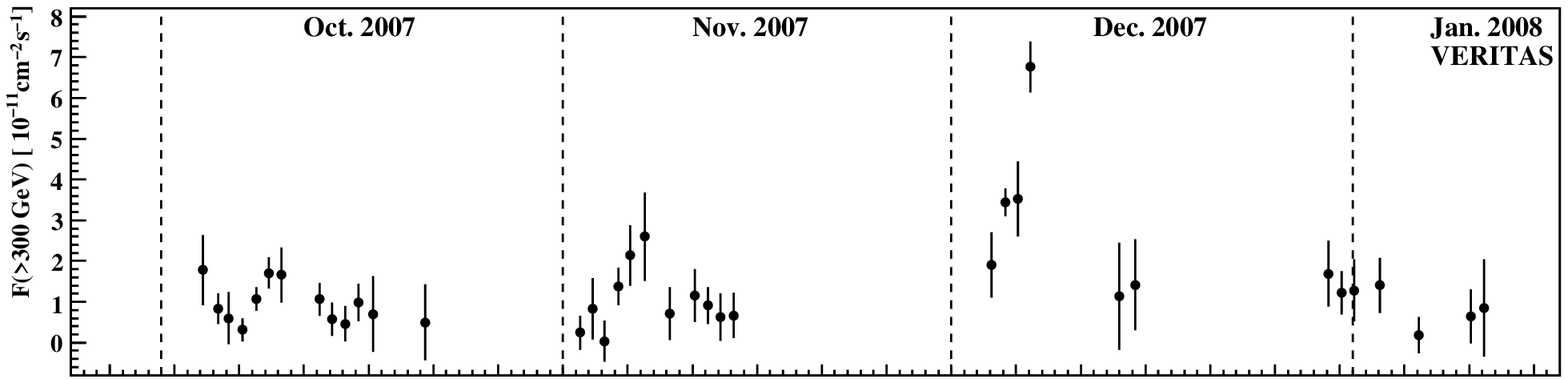}
\includegraphics[width=5.9in]{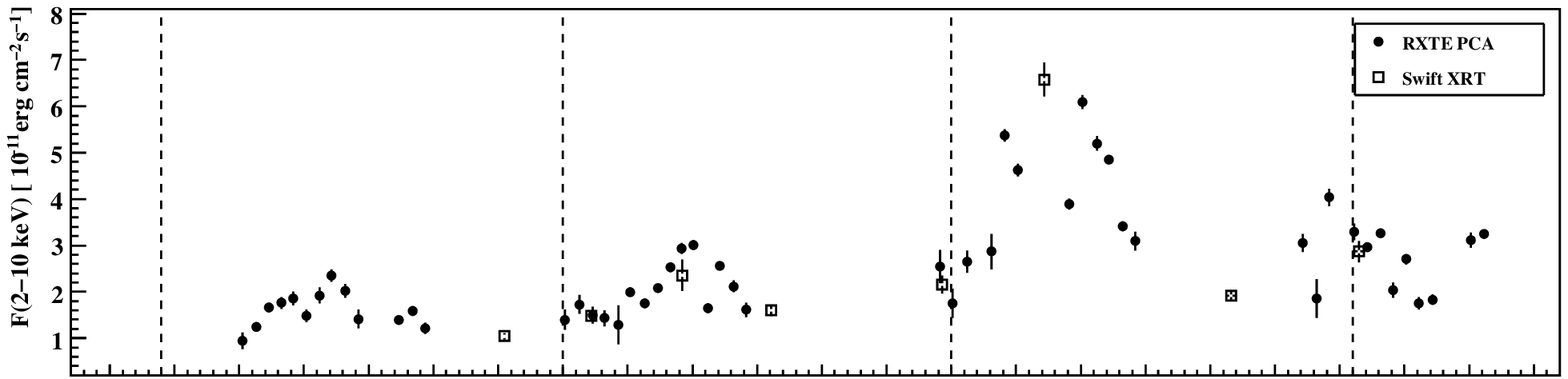}
\includegraphics[width=5.9in]{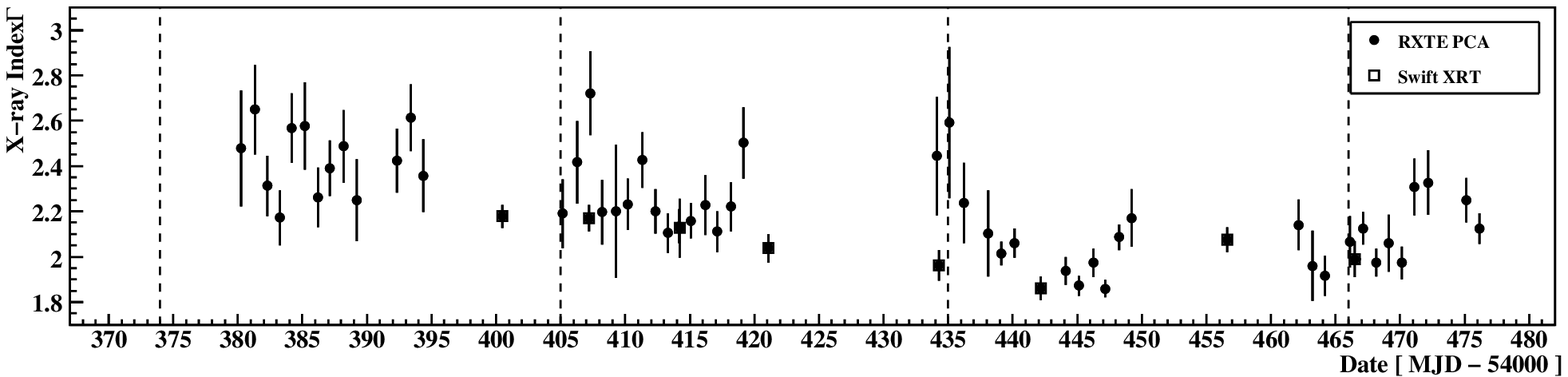}
\caption{VHE $\gamma$-ray and X-ray light curve of 1ES\,2344+514. Shown 
in the top panel are the nightly VERITAS F($>$300 GeV) fluxes. The middle 
panel shows the 2--10 keV fluxes from observations with \emph{RXTE} PCA 
(circles) and \emph{Swift} XRT (open squares). The bottom panel shows the 
power law indices $\Gamma$ from the 0.4--10\,keV \emph{Swift} XRT spectra 
and 3--20\,keV \emph{RXTE} PCA spectra.} 
\label{Fig2344lc}
\end{figure*}
\begin{figure}
\centering
\includegraphics[width=3.0in]{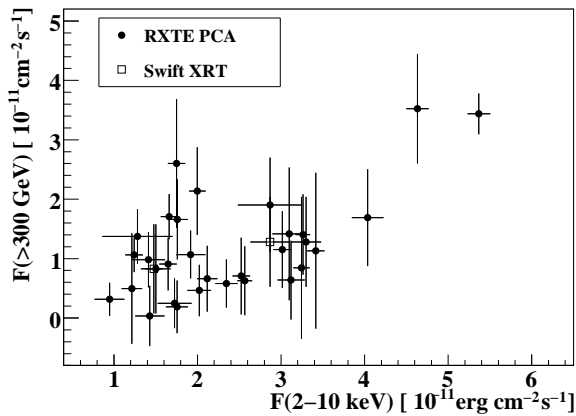}
\caption{VERITAS $\gamma$-ray flux F($>$300 GeV) versus X-ray 2--10 keV flux 
for 1ES\,2344+514 observations from nights with \emph{RXTE} PCA (circles) 
and \emph{Swift} XRT (open squares) data.}
\label{Fig2344Fcorr}
\end{figure}
1ES\,2344+514 is another close HBL (z $=$ 0.044), and was first detected 
at VHE $\gamma$-ray energies by the Whipple 10 m telescope in 1995 
\cite{Catanese98}. Figure \ref{Fig2344lc} shows the nightly 
VHE $\gamma$-ray and X-ray light curve of 1ES\,2344+514 from VERITAS, 
\emph{RXTE} PCA, and \emph{Swift} XRT observations. A strong 
VHE $\gamma$-ray flare is seen on December 7, 2007 (54441.12 MJD) 
at a flux F($>$300 GeV) corresponding to 48\% of the Crab Nebula flux. 
The measured increase in flux of a factor of 1.92 $\pm$ 0.53 between the 
previous night and the flare night shows the first clear evidence 
of $\sim$day time-scale VHE $\gamma$-ray variability from 1ES\,2344+514 
since the initial Whipple 10 m detection in 1995. Excluding the December 
7 flaring event, the average F(E$>$300 GeV) is 7.6\% of the Crab Nebula flux. 
For the full VERITAS data set a high level of variability 
F$_{\rm{var}} = (75 \pm 10$)\% is implied. Excluding the flare night, 
a F$_{\rm{var}} = (34 \pm 16$)\% is determined. 

Figure \ref{Fig2344lc} (lower panels) shows the 2--10 keV flux and photon 
index $\Gamma$ measured over 3--20 keV from \emph{RXTE} PCA and 0.4--10 keV 
from \emph{Swift} XRT data. The X-ray flux is shown to be highly variable 
throughout the campaign, with F$_{\rm{var}} = (51 \pm 1$)\%. In December 2007, 
large amplitude flaring is evident with flux doubling time-scales of $\sim$1 
day. A 2--10 keV flux of $(6.58 \pm 0.37) \times 10^{-11}$ erg cm$^{-2}$ 
s$^{-1}$ is seen from the Swift XRT data on December 8, 2007, representing 
the highest X-ray flux ever measured for 1ES\,2344+514. Figure 
\ref{Fig2344Fcorr} shows the VERITAS flux F($>$300 GeV) versus \emph{RXTE} 
PCA and \emph{Swift} XRT 2--10 keV fluxes for nights with observations in both 
energy bands. A Pearson coefficient of $r = 0.60 \pm 0.11$ is calculated for 
the VHE $\gamma$-ray to X-ray flux points, suggestive of 
correlated variability. Further results and discussion on this campaign are 
in press \cite{Acciari09}. 
\section{Conclusions}  
Joint VHE $\gamma$-ray and X-ray observing campaigns on the bright 
HBLs Mrk\,421, Mrk\,501, and 1ES\,2344+514 with VERITAS reveal significant 
flux variability on nightly time-scales. X-ray spectral hardening is shown 
at increasing flux levels. Results from synchrotron self-Compton (SSC) 
modeling of the broadband SED will be presented at the conference.
\section*{Acknowledgments}
{\footnotesize 
This research is supported by grants from the US Department of Energy, 
the US National Science Foundation, and the Smithsonian Institution, 
by NSERC in Canada, by Science Foundation Ireland, and by STFC in the UK. 
We acknowledge the excellent work of the technical support staff at the 
FLWO and the collaborating institutions in the construction and operation 
of the instrument.}

\end{document}